\documentclass[floats,aps,prb,draft,preprint,showpacs]{revtex4}

\usepackage{amssymb}

\usepackage{graphics}
\usepackage{graphicx}
\usepackage{epsfig}
\usepackage{amssymb}

\usepackage{amsmath}

\begin{document}



 \title{Stabilization of the surface CDW order parameter by long-range Coulomb
interaction
}
\author{P. P. Aseev}
\author{S. N. Artemenko}\email{art@cplire.ru}
 \affiliation{Institute for Radio-engineering and Electronics of
Russian Academy of Sciences \\
Mokhovaya 11-7, Moscow 125009, Russia}
\affiliation{Moscow Institute for Physics and Technology\\
Institutski
per. 9, Dolgoprudny, Russia}
\date{\today}

\begin{abstract}
We study theoretically formation of two-dimensional (2D) charge density wave (CDW)
in a system of conducting chains at the surface of an insulator due to
interaction of quasi 1D surface electrons with phonons. We show that the
unscreened long-range Coloumb interaction between the charges induced by
fluctuations of the CDW phase stabilizes the finite order parameter
value at finite temperatures, and thus the long-range order (LRO) exists.
In the case of screened Coloumb interaction the phase fluctuations
suppress the phase transition, but decay of the order parameter is
rather slow, it obeys a power-law $\langle \Delta^*(r) \Delta(0)
\rangle\propto r^{-\gamma}$ with small exponent $\gamma$.
\end{abstract}

\maketitle

In contrast to three-dimensional (3D) systems where fluctuations are usually small, in
2D systems fluctuations can greatly affect behaviour of a system. It is
well-known~\cite{Mermin-Wagner} that in low-dimensional (1D and 2D)
continuous systems with sufficiently short-range interaction a
long-range order (LRO) is suppressed by long-range fluctuations. These fluctuations can
be excited with little energy cost and they are favored since they increase the entropy. Thus, the 2D charge-density wave (CDW) phase is also believed to be suppressed by fluctuations of the  CDW phase.
However, there are experimental evidences of existence of 2D CDW in layered~\cite{Marynowski} and linear-chain compounds~\cite{Monceau}. In the latter case it was shown that the
critical temperature for a surface CDW in $NbSe_3$ is higher than for a bulk one.

In a 2D superconductor which is related system, it is known~\cite{VarlamovLarkin} that there is no
LRO indeed. Below the Berezinski-Kosterlitz-Thouless critical temperature, the correlation function of the order
parameter obeys a power-law resulting in a pseudo-LRO in the system.

In this paper we examine the decay of the order parameter of the incommensurate CDW at the
surface (commensurability is expected to stabilize the LRO). 

The Mermin-Wagner theorem is applied~\cite{Mermin-Wagner} to the case
of sufficiently short-range interaction. One of our goals is to study how the
long-range Coulomb interaction may affect the LRO of the CDW. We
consider three models of electron-electron interaction: non-screened
Coloumb interaction, an interaction between surface electrons screened
by electrons in the bulk, and an interaction between surface electrons
screened only by surface electrons. The first case can be applied when the size
of a sample is not greater than the screening radius. The second case refers
to the systems where the bulk material is a semiconductor and there are electrons in the bulk or to systems with a gate.
Finally, the third case refers to the systems in which the bulk material is an
insulator and the only screening electrons are the electrons thermally
excited over the Peierls gap of the 2D CDW.

Below we set $\hbar$ and $k_{\mathrm{B}}$ to unity, restoring dimensional units in
final expressions when necessary.

We study a system of conducting chains at the surface of a 3D
insulator, electron system at the surface being considered as a 2D
electron gas interacting with acoustic phonons in the
bulk. We assume that $x$-axis is taken along the nesting vector $Q$.
$y$-axis is taken perpendicular to the nesting vector and parallel
to the surface. Finally, $z$-axis is the axis normal to the surface.

First, we consider the case of non-interacting electrons. The total
action $S$ is the sum of an action of free 2D electrons $S_{\mathrm{e}}$, action of
free phonons in the bulk $S_{\mathrm{ph}}$, and action of electron-phonon
interaction $S_{\mathrm{e-ph}}$
\begin{equation}
	S = S_{\mathrm{e}} + S_{\mathrm{ph}} + S_{\mathrm{e-ph}}
	\label{eqn:action}
\end{equation}
Only the phonons with the  wave vector component along the surface
(longitudinal component) close to the nesting vector $\pm Q$
are relevant for formation of the CDW. If one does not take into account fluctuations of
phonon modes one should consider only the modes with
longitudinal component of wave vector equal to $\pm Q$ exactly and the corresponding
creation-annihilation operators $\hat b_{\pm Q}$, $\hat
b^\dag_{\pm Q}$. However, since we take into account the fluctuations we
allow the phonons to have the longitudinal component of the wave
vector that slightly differs from $Q$. We describe these phonons with
components of wave-vector $\pm Q + q_x$, $q_y$ and $q_z$ (where
$q_x \ll Q$) using
creation-annihilation operators $\hat b_{\pm Q}(q_x,q_y,q_z)$. The phonon field
$\hat \varphi$ is introduced in a standard way
\begin{equation}
		\hat\varphi_{Q}({q},\omega) =
		\sqrt{\frac{\omega_0
		(q)}{2}} \left(\hat b_{Q}(
		q,\omega)+ \hat b^\dag_{-Q}(-q,-\omega)  \right)
	\label{eqn:phonon-field},
\end{equation}
where $\omega_0(q) = s\sqrt{(Q+q_x)^2+q_y^2+q_z^2}$ is the spectrum of
acoustic phonons. Then the Matsubara action of free phonons reads
\begin{equation}
	S_{\mathrm{ph}} = -T\sum\limits_{\omega} \int \frac{d^3 q}{(2\pi)^3}
	\frac{\omega^2+\omega_0^2(q)}{\omega^2_0(q)}
	|\varphi_Q(q, \omega)|^2
	\label{eqn:S-ph},
\end{equation}
where $T$ is temperature, and $q$ stands for
all the three components $q_x$, $q_y$, $q_z$

Since electrons are confined in the direction normal to the surface the
electron field operator can be written as
\begin{equation*}
\hat \Psi(x,y,z) = \hat\Psi(x,y) w(z)
\end{equation*}
where $w(z)$ is the wave function of the ground mode of size quantization, $w(z)$ decays
exponentially in the depth of the bulk material:
\begin{equation}
	W(z)=|w(x,y,z)|^2 = \kappa e^{-\kappa z}
	\label{eqn:wz-kappa}
\end{equation}
We do not take into account higher transversal modes because their
contribution is exponentially small at low temperatures.
Besides, the low-temperature behaviour of the system and particularly the
properties of CDW are determined by the electrons with energy close to
the Fermi energy, and, therefore, with momentum close to $\pm Q/2$. Thus we can represent the electron field as
\begin{equation*}
	\hat \Psi(x,y) = e^{iQx/2}\hat \psi_+(x,y) + e^{-iQx/2}\hat
	\psi_-(x,y),
\end{equation*}
where $\hat\psi_{\pm}(x,y)$ vary smoothly in comparison with the
correspondent exponential factor. The Matsubara action of free electrons
reads
\begin{equation}
	S_{\mathrm{e}} =
		\int \frac{d^2 k}{(2\pi)^2} d\tau\left\{ \psi_+^*\left[\partial_\tau -
		\varepsilon_+\right]\psi_+ + \psi_-^* \left[
		\partial_\tau - \varepsilon_-
		\right]\psi_-
		\right\},
		\label{eqn:S-e}
\end{equation}
	where $\varepsilon_{\pm} = \varepsilon_0(k \pm Q/2)$ and
	$\varepsilon_0(k)$ is a spectrum of free electrons.
	
The term describing the electron-phonon interaction reads
\begin{multline}
	S_{\mathrm{e-ph}} =
	- \int d^3r'd^3r d\tau \psi_+^*(r)
	\psi_-(r)\times\\\times W(z)g(r-r')\varphi_Q(r') + c.c.
	\label{eqn:S-eph-1}
\end{multline}
We assume that the Fourier transform of interaction potential $g(k) =
\int d^3 r g(r) e^{-ikr}$ may depend on the momentum $k$. Although we do not
specify the particular dependence, note that  such a dependence
appears if one considers an electron-phonon interaction in a system
with anisotropy.

It is convenient to introduce the order parameter as follows:
\begin{equation}
	\Delta (k_x,k_y) =  \int \frac{d q_z}{2\pi}
	\varphi_{Q,q_z}(k_x,k_y)g_{q_z}(k_x,k_y)W_{q_z}
	\label{eqn:order}
\end{equation}
where $W_{q_z}$ is the Fourier transform of the square of transversal
wavefunction~(\ref{eqn:wz-kappa}) $W_{q_z} = \int dz W(z) e^{-iq_z z}$. The
electron-phonon contribution to action~(\ref{eqn:S-eph-1}) can be rewritten in terms of
$\Delta$ in a quite simple form
\begin{equation}
	S_{\mathrm{e-ph}} = - \int d^2 rd\tau \left\{\psi_+^*\psi_- \Delta
		+ \Delta^* \psi_-^* \psi_+ \right\}
	\label{eqn:S-e-ph2}
\end{equation}

If one does not take into account fluctuations of the phonon modes then
$\Delta$ does not depend on coordinates and can be taken real. In this case one can calculate
electron Green functions in imaginary time
\begin{multline*}
	G_{\pm\pm} = \langle \psi^\dag_{\pm} \psi_{\pm} \rangle
	=\\= \int \mathcal{D}\psi^*_{+}\mathcal{D}\psi_+
	\mathcal{D}\psi^*_{-}\mathcal{D}\psi_-\;\psi^*_{\pm}\psi_{\pm}
	e^{-S_{\mathrm{e}}-S_{\mathrm{e-ph}}}
	\label{}
\end{multline*}
and obtain the following expressions
\begin{align*}
			G_{+-}(\varepsilon,k) &= G_{-+}(\varepsilon,k) =
			-\frac{\Delta}{\varepsilon^2+\xi^2+\Delta^2}\\
			G_{++}(\varepsilon,k) &= - G_{--}^* =
			-\frac{-i\varepsilon+\xi}{\varepsilon^2+\xi^2+\Delta^2}
\end{align*}
where $\xi = \frac{\varepsilon_+(k)-\varepsilon_-(k)}{2}\approx
v_{\mathrm{F}}k_x$.
These Green functions correspond to an electron spectrum with a
gap $\varepsilon = \sqrt{\xi^2+|\Delta|^2}$. This result is similar to
3D case \cite{CDW-3D}.

It is convenient to integrate out the phonon fields and thus to
derive an effective action for $\Delta$. Using the definition for
$\Delta$~(\ref{eqn:order}) we obtain
\begin{multline*}
	\int \mathcal{D}\varphi^*_Q\mathcal{D}\varphi_Q
	e^{-S_{\mathrm{ph}}} =
	\int
	\mathcal{D}\varphi^*_Q\mathcal{D}\varphi_Q\mathcal{D}\Delta^*\mathcal{D}\Delta\mathcal{D}\beta^*\mathcal{D\beta}\times\\\times
	\exp\left\{-S_{\mathrm{ph}}+\beta^*(k_x,k_y)\left[ \Delta(k_x,k_y)
	-\right.\right.\\-\left.\left. \int
	\varphi^*_{Q,q_z}(k_x,k_y)g^*_{q_z}(k_x,k_y)W^*_{q_z}
	\frac{d q_z}{2\pi}\right] +
	c.c.
	\right\}
	\label{}
\end{multline*}
where $\beta(k_x,k_y)$, $\beta^*(k_x,k_y)$ are Lagrange multipliers for
$\Delta^*$ and $\Delta$. Performing Gaussian integration over
$\varphi$,$\varphi^*$,$\beta$,$\beta^*$ we obtain an effective action
for $\Delta$
\begin{align}
	S_{\Delta} =& -T \sum\limits_{\omega}\int \frac{dk_x dk_y}{(2\pi)^2}
	|\Delta(\omega,k)|^2/F(\omega,k) \label{eqn:S-ph2}\\
	F(\omega,k) =& \int \frac{dq_z}{2\pi}
	\frac{\omega_0^2(k,q_z)}{\omega^2+\omega_0^2(k,q_z)}|g_{q_z}^2(k_x,k_y)|
	|W^2_{q_z}|
\end{align}
We can expand $F(\omega, k)$ at small $\omega$ and $k$
$$
F(\omega,k) = \frac{g_0^2 \kappa}{2} \left[1- \frac{\omega^2 +
		s'^2 k^2}{s^2(Q^2-\kappa^2)}  \right]
$$
where $g_0=g(k=0)$, and $s'$ can be roughly estimated as 
\begin{equation}
s'^2 \approx s^2
\frac{|g(Q)-g_0|}{g_0}
\label{eqn:s}
\end{equation}
We consider the case when $Q>\kappa$. If $Q<\kappa$ the formation of CDW
due to interaction with the phonons in the bulk is impossible.

Now we can minimize the total action given
by~(\ref{eqn:action}),~(\ref{eqn:S-e}),~(\ref{eqn:S-e-ph2}),~(\ref{eqn:S-ph2})
and thus find the classical solution for the order parameter $\Delta$
(i.e. the solution that does not take into account fluctuations of
phonon field and, therefore, fluctuations of the order parameter). The
equation for the  classical $\Delta$ resembles the self-consistent condition for
the gap in the BCS theory	
\begin{equation*}
			1 = 
			\lambda \int
	\frac{1}{\sqrt{\varepsilon^2-|\Delta|^2}}\tanh
	\frac{\varepsilon}{2T}d\varepsilon,
\end{equation*}
where $\lambda = \frac{g_0^2 \kappa}{4\pi v_{\mathrm{F}} a_y}$, $a_y$ is a lattice
constant in the direction normal to the nesting vector. The cut-off
parameter at high energies is of order of $\varepsilon_F$ and thus the solution
at $T=0$ is 
$
\Delta = \varepsilon_F e^{-\lambda}.
$
The result is similar to the case of CDW in 3D. The main
feature of 2D systems is that in constrast to 3D, fluctuations are
important in 2D. The main contribution to the fluctuations of the order
parameter is given by fluctuations of the phase $\chi$ where
$\Delta=\Delta_0 e^{i\chi}$. The amplitude of the order parameter determines the Peierls energy gap, and fluctuations of the ampltitude are described by the mode with a gap in the spectrum, in contrast to the long wavelength fluctuations of the
phase which are related to gapless modes, so that at $q \to 0$ they do not affect the total energy. Thus at $T \ll \Delta$ we can ignore amplitude fluctuations, and it is convenient to
rewrite the total action in terms of $\chi$
Then the phonon action reads
\begin{equation*}
	S_{\Delta} = -T|\Delta_0|^2\sum_{\omega}\int \frac{d^2 q}{(2\pi)^2}
	\frac{\chi(\omega,q)\chi(-\omega,-q)}{F^2(\omega,q)}
\end{equation*}

The total effective action is
\begin{equation}
	S_{\mathrm{eff}}[\chi^*,\chi] = S_{\Delta} - \ln \int
	\mathcal{D}\psi_+^*\mathcal{D}\psi_+\mathcal{D}\psi_-^*\mathcal{D}\psi_-
	e^{-S_{\mathrm{e}} - S_{\mathrm{e-ph}}}
	\label{}
\end{equation}
Expanding the exponent in powers of $\chi$
and leaving the second order terms, we perform Gaussian integration
over $\psi^*_{\pm}$, $\psi_{\pm}$. After some algebra we obtain the
following effective action
\begin{multline*}
	S_{\mathrm{eff}} = 
	T\sum\limits_{\omega}\int \frac{d^2k}{(2\pi)^2}
	\frac{2|\Delta|^2}{g_0^2 \kappa} \times\\\times \left\{\left[
				\frac{\lambda}{2} +
				\frac{\omega^2+s'^2|k|^2}{s^2(Q^2+\kappa^2)}+
				\frac{\lambda}{6}\frac{\omega^2+v_{\mathrm{F}}^2k_x^2}{|\Delta|^2}
				\right]\chi(\varepsilon,k)\chi(\varepsilon,k)
				+\right.\\+\left.
				\left[-\frac{\lambda}{2}+\frac{\lambda}{12}\frac{\omega^2+v_{\mathrm{F}}^2
				k_x^2}{|\Delta_0|^2}
				\right]\chi(\varepsilon,k)\chi(-\varepsilon,-k)
				\right\}
\end{multline*}
Green function for $\chi$ can be easily calculated
\begin{align}
	\langle \chi(\omega,k) \chi(-\omega,-k) \rangle
	&= \int \mathcal{D} \chi \; \chi^2(\omega,k) e^{-S[\chi]}
	\label{eqn:green-chi}\\
			\left\langle \chi(\omega, k) \chi(\omega,
			k)\right\rangle
			&=
			\frac{\pi\lambda
			v_{x}^2/(8v_{\mathrm{F}})}{\omega^2+v^2_{x}k_x^2
			+ s'^2 k_y^2}
\end{align}
where $v^2_{x} \simeq v^2_F
		\frac{s^2 \left(Q^2-\kappa^2  \right)}{|\Delta|^2}$
		is a velocity of excitations along the $x$-axis. Note
		that it follows from~(\ref{eqn:s}) that the velocity of excitations $s'$ along the $y$-axis is
		non-zero if we take into account a dispersion of
		electron-phonon interaction $g(k)$. However, one would
		obtain non-zero velocity of excitations along the
		$y$-axis if one took into account non-ideal nesting
		conditions. Thus we can assume that in a real setup the
		value of $s'$ is of order of sound velocity $s$.
		The correlation function for phase $\chi$ can be
		calculated
		as
		\begin{multline}
			\left\langle \chi(r,t) \chi(r+\delta r,t) \right\rangle
			=\\=
			\sum\limits_{\omega}\int \frac{d^2k }{(2\pi)^2}
			\left\langle \chi(\omega,k) \chi(-\omega,-k)
			\right\rangle e^{ik \delta r} 
			= \\=
			\frac{\pi \lambda}{32 v_{\mathrm{F}}}\int
			\frac{d^2 k}{(2\pi)^2}\frac{e^{ik\delta
			r}}{\sqrt{v_x^2k_x^2+s'^2k_y^2}}\coth
			\frac{\sqrt{v_x^2k_x^2 + s'^2k_y^2}}{2T}
			\label{eqn:integral-nocoulomb}
		\end{multline}
If the temperature $T=0$ then the correlation function for phase is finite
and therefore the LRO exists in the system. Otherwise,
if the temperature $T>0$, the integral in~(\ref{eqn:integral-nocoulomb})
diverges at small $k$ when $\delta r=0$. Thus there is no long-range
order at finite temperatures. However, the correlation function for the order
parameter $\langle \Delta^*(r,t) \Delta(r+\delta r,t) = |\Delta_0|^2
e^{-\langle [\chi(\delta r,t)-\chi(0,t)] \chi(0,t)   \rangle}$ obeys the
power-law	
\begin{equation*}
\langle \Delta^*(0) \Delta(r) \rangle 
=
		|\Delta_0|^2
		\left(r Q  \right)^{-\gamma_0},\;\\ \gamma_0 =
		\frac{\lambda}{8}
		\frac{v_{x}}{v_{\mathrm{F}}}\frac{T
		a_y}{\hbar s'}
\end{equation*}
The exponent $\gamma_0$ is of order of $\frac{T}{\Theta}$, where $\Theta$
is the Debye temperature.

In the presence of Coulomb interaction there is an addition contribution to the total
energy, and thus to the total action~(\ref{eqn:action}), due to an
interaction of charges induced by phase fluctuations. Given the
phase $\chi$, the electron density can be calculated as
\begin{equation*}
	\rho = \int
	\mathcal{D}\psi_+^*\mathcal{D}\psi_+\mathcal{D}\psi_-^*\mathcal{D}\psi_-\;
	\left(\psi_+^*\psi_+ + \psi_-^*\psi_-  \right)e^{-S}
	\label{}
\end{equation*}
Thus we obtain the electron density induced by phase fluctuations
	$\delta \rho = - e \partial_x \chi/(\pi a_y)$
.

The corresponding term in the action responsible for Coulomb interaction
is given by
\begin{equation*}
	S_{\mathrm{C}} = \int d\tau d^2r_1 d^2 r_2
		\delta\rho(r_1) V(r_1-r_2) \delta\rho(r_2)
	\label{}
\end{equation*}
We consider three different models for Coulomb potential $V(r_1-r_2)$:
non-screened Coloumb potential, Coloumb potential screened by free
electrons in 3D material, and Coloumb potential screened by  thermally excited
electrons in 2D electron layer.

First we consider non-screened Coulomb potential $V(r)=e^2/(\epsilon^*
r)$, where $\epsilon^*$ is effective dielectric constant. If the surface
is a boundary between two media with dielectric constants $\epsilon_1$
and $\epsilon_2$, then $\epsilon^*=(\epsilon_1+\epsilon_2)/2$
Calculating the Green function for $\chi$ by using~(\ref{eqn:green-chi})
we obtain
\begin{equation*}
	\left\langle \chi(\omega,q) \chi(-\omega,-q) \right\rangle
	= \frac{\lambda\pi v_{x}^2/(8v_{\mathrm{F}})}{\omega^2 +
	\frac{2\lambda e^2}{\epsilon^*a_y}\frac{v_{x}^2}{v_{\mathrm{F}}}
		q_x+  s'^2q_y^2}
\end{equation*}
The singularity at low frequencies is now integrable if $T>0$ as well, so the mean square
of fluctuations of phase 
$\left\langle \chi(r,t) \chi(r,t) \right\rangle$ is finite and therefore the LRO
exists in the system.

Consider Coulomb interaction screened by free electrons in 3D, for
example by electrons in a gate or by conduction electrons in a semiconductor.
The interaction is described by Yukawa potential
$V(r)=\frac{e^2}{\epsilon^*r}e^{-r/r_{\mathrm{D}}}$, where $r_{\mathrm{D}}$ is a screening radius. The
Fourier transformation of this potential reads
\begin{equation*}
	V(q) = \int d^2 r V(r) e^{-iqr}  =2\pi e^2/\left(\epsilon^*\sqrt{q^2 +
	r_{\mathrm{D}}^{-2}}\right),
\end{equation*}
and the Green function for $\chi$ calculated using~(\ref{eqn:green-chi}) has the form
$$
		\left\langle \chi(\omega,q) \chi(-\omega,-q) \right\rangle =
$$
$$
		= \frac{\lambda\pi v_{x}^2/(8v_{\mathrm{F}})}{\omega^2 +v_{x}^2\left(1+
		12\lambda\frac{e^2}{\epsilon^*v_{\mathrm{F}}}\frac{r_{\mathrm{D}}}{a_y}
	 \right)q_x^2+  s'^2q_y^2}
$$
The integral over $\omega$ and $q$ diverges at $T>0$ as in the case when there is
no inter-electronic interaction, thus there is no LRO.
However, the correlation function for $\Delta$ obeys a power-law with different exponent.
$$
	\langle \Delta^*(0) \Delta(r) \rangle = \Delta_0 \left(
	\frac{r}{r_{\mathrm{D}}} \right)^{-\gamma_3},
$$
$$
\gamma_3 =
		\gamma_0
		\sqrt{\frac{12\epsilon^*\lambda \hbar
		v_{\mathrm{F}}}{e^2}}\sqrt{\frac{a_y}{r_{\mathrm{D}}}}
$$
The exponent $\gamma_3\ll \gamma_0$, and although there is only a
short-range order in the system, the correlations decay more slowly and the system
behaves like it has a real LRO provided the screening radius is large enough.

Finally, we consider the Coulomb interaction screened only by electrons thermally excited over the Peierls gap
in the conducting 2D layer of width $a_z$. The Fourier transformation
of the potential is given by
\begin{equation*}
	V(q)=2\pi e^2/\left[\epsilon^*\left(|q|+ a_z/r_\mathrm{D}^2\right)\right]
	\label{}
\end{equation*}
Density of screening electrons and, therefore, the screening radius
$r_\mathrm{D}$ depends exponentially on temperature
$r^2_{\mathrm{D}} \propto
n_e^{-1} \propto e^{\Delta/T}$.

The Green function for $\chi$ calculated using~(\ref{eqn:green-chi}) reads
\begin{multline*}
	\left\langle \chi(\omega,q) \chi(-\omega,-q) \right\rangle
	=\\= \frac{\lambda\pi v_{x}^2/(8v_{\mathrm{F}})}{\omega^2 +v_{x}^2\left(1+
	12\lambda\frac{e^2}{\epsilon^*v_\mathrm{F}}\frac{r^2_D}{a_ya_z}
	 \right)q_x^2+  s'^2q_y^2}
	\label{}
\end{multline*}
The integral over $\omega$ and $q$ diverges at $T>0$ again, and the
correlation function for $\Delta$ obeys a power-law but with a smaller
exponent
\begin{multline*}
	\langle \Delta^*(0) \Delta(r) \rangle = \Delta_0 \left(
	ra_z/r_{\mathrm{D}}^2 \right)^{-\gamma_2},
\\
\gamma_2 =
	 	\frac{\sqrt{\lambda}}{16\sqrt{3}}\frac{v_{x}}{s'}\frac{Ta_y}{\hbar
		v_{\mathrm{F}}}\sqrt{\frac{\epsilon^*\hbar
		v_{\mathrm{F}}}{e^2}}\sqrt{\frac{a_ya_z}{r^2_D}}
\label{}
\end{multline*}
The exponent $\gamma_2 \ll \gamma_3 \ll \gamma_0$ for similar values of screening radius, and the correlations
decay at greater distance than in the case of 3D screening or
non-interacting electrons. Strictly saying, there is no LRO
in the system but due to the very slow power-law  decay of correlations there
is a pseudo-LRO.

The work was supported by Russian Foundation for Basic Research and
Russian Ministry of Education and Science (grant No 16.513.11.306).



\end{document}